\documentclass[twocolumn]{aastex63}

\usepackage{xspace}

\newcommand{\todo}{\ifmmode \text{\color{purple}\Huge{\(\bullet\)}} \else {\color{purple}{\Huge$\bullet$}}\fi}
\newcommand{\finish}{\ifmmode \text{\color{blue}\Huge{\(\bullet\)}} \else {\color{blue}{\Huge$\bullet$}}\fi}

\newcommand{\target}{BBQSORS\xspace}

\newcommand{\mgtwo}{Mg\,\textsc{ii}$\lambda2800$\xspace}
\newcommand{\aliiidou}{Al\,\textsc{iii}$\lambda\lambda1855,1863$\xspace}
\newcommand{\mgtwodou}{Mg\,\textsc{ii}$\lambda\lambda2796,2803$\xspace}
\newcommand{\ciii}{C\,\textsc{iii}{]}$\lambda1909$\xspace}
\newcommand{\feii}{Fe\,\textsc{ii}\xspace}
\newcommand{\oi}{O\,\textsc{i}$\lambda8446$\xspace}
\newcommand{\halpha}{\mathrm{H\alpha}\xspace}
\newcommand{\lya}{\mathrm{Ly\alpha}\xspace}
\newcommand{\NH}{N_\mathrm{H}}

\newcommand{\msun}{M_{\odot}}

\newcommand{\lfirst}{L_\mathrm{1.4GHz}}

\newcommand{\av}{A_\mathrm{V}}
\newcommand{\R}{\mathcal{R_{\rm rest}}}

\newcommand{\sanfit}{\mathtt{S^{3}Fit}}
\newcommand{\tnr}[2]{{#1}_\mathrm{#2}}

\newcommand{\kms}{\mathrm{km\ s^{-1}}}
\newcommand{\ergs}{\mathrm{erg\ s^{-1}}}
\newcommand{\whz}{\mathrm{W}~\mathrm{Hz}^{-1}}

\newcommand{\mujy}{\mathrm{\mu Jy}}
\definecolor{darkgoldenrod}{rgb}{0.72, 0.53, 0.04}

\usepackage{natbib}
\usepackage{amsmath}
\usepackage{multirow} %for using \multirow
\usepackage{ifthen} %if/then package
\usepackage{txfonts}
\usepackage{tabularx}
\usepackage{CJKutf8}
\received{\today}
\usepackage{lineno}

\submitjournal{ApJL}

\shorttitle{BBQSORS}
\shortauthors{Zhong et al.}
\begin{document}
\begin{CJK*}{UTF8}{gkai}
% \title{Discovery of a Quasar at $z=1.715$ with Blackbody Continuum of $T\sim10^4$~K: Transitioning from an LRD to a Normal Quasar?}

\title{Blackbody Quasar and Radio Source (BBQSORS): A Candidate of Transitional Little Red Dots with a $T\sim10^4$~K Blackbody Spectrum}

\correspondingauthor{Yuxing Zhong, Xiaoyang Chen, Kohei Ichikawa}
\email{yuxing.zhong.astro@gmail.com, astro@xychen.me, k.ichikawa@astr.tohoku.ac.jp}

\author[0009-0001-3910-2288]{Yuxing Zhong (仲宇星)}
\affiliation{Frontier Research Institute for Interdisciplinary Sciences, Tohoku University, Sendai, Miyagi 980-8578, Japan}
\affiliation{Department of Physics, School of Advanced Science and Engineering, Faculty of Science and Engineering, Waseda University, 3-4-1, Okubo, Shinjuku, Tokyo 169-8555, Japan}
%\email[show]{yuxing.zhong.astro@gmail.com}

\author[0000-0003-2682-473X]{Xiaoyang Chen}
\affiliation{Frontier Research Institute for Interdisciplinary Sciences, Tohoku University, Sendai, Miyagi 980-8578, Japan}
%\email[show]{xy.chen@astr.tohoku.ac.jp}

\author[0000-0002-4377-903X]{Kohei Ichikawa}
\affiliation{Frontier Research Institute for Interdisciplinary Sciences, Tohoku University, Sendai, Miyagi 980-8578, Japan}
\affiliation{Astronomical Institute, Tohoku University, Aramaki, Aoba-ku, Sendai, Miyagi 980-8578, Japan}

\author[0009-0002-4201-7727]{Youwen Kong}
\affiliation{Institute of Astronomy, Graduate School of Science, The University of Tokyo, 2-21-1 Osawa, Mitaka, Tokyo, 181-0015 Japan}

\author[0000-0003-4569-1098]{Kentaro Aoki}
\affiliation{Subaru Telescope, National Astronomical Observatory of Japan, 650 North A'ohoku Place, Hilo, HI 96720, U.S.A.}

\author[0000-0002-9754-3081]{Satoshi Yamada}
\affiliation{Frontier Research Institute for Interdisciplinary Sciences, Tohoku University, Sendai, Miyagi 980-8578, Japan}
\affiliation{Astronomical Institute, Tohoku University, Aramaki, Aoba-ku, Sendai, Miyagi 980-8578, Japan}
\affiliation{Department of Astronomy, University of Geneva, Ch.d’Ecogia 16, 1290, Versoix, Switzerland.}

\author[0000-0002-7402-5441]{Tohru Nagao}
\affiliation{Research Center for Space and Cosmic Evolution, Ehime University, 2-5 Bunkyo-cho, Matsuyama, Ehime 790-8577, Japan}
\affiliation{Amanogawa Galaxy Astronomy Research Center, Kagoshima University, 1-21-35 Korimoto, Kagoshima 890-0065, Japan}

\author[0009-0000-0538-6780]{Daisaburo Kido}
\affiliation{Research Center for the Early Universe, Graduate School of Science, The University of Tokyo, Bunkyo, Tokyo}

%Alphabetical below

\author[0000-0002-3866-9645]{Toshihiro Kawaguchi}
\affiliation{Graduate School of Science and Engineering, University of Toyama, Gofuku 3190, Toyama 930-8555, Japan}

\author[0000-0001-5063-0340]{Yoshiki Matsuoka}
\affiliation{Research Center for Space and Cosmic Evolution, Ehime University, 2-5 Bunkyo-cho, Matsuyama, Ehime 790-8577, Japan}

\author[0000-0002-5464-9943]{Toru Misawa}
\affiliation{Center for General Education, Shinshu University, 3-1-1 Asahi, Matsumoto, Nagano 390-8621, Japan;}

\author[0000-0002-6068-8949]{Shoichiro Mizukoshi}
\affiliation{Academia Sinica Institute of Astronomy and Astrophysics, 11F of Astronomy-Mathematics Building, AS/NTU, No.1, Section 4, Roosevelt Road, Taipei 10617, Taiwan}

\author[0000-0003-2984-6803]{Masafusa Onoue}
\affiliation{Waseda Institute for Advanced Study (WIAS), Waseda University, 1-21-1, Nishi-Waseda, Shinjuku, Tokyo 169-0051, Japan}

\author[0000-0003-3769-6630]{Ayumi Takahashi}
\affiliation{Faculty of Science, Kanagawa University, 3-27-1 Rokkakubashi, Kanagawa-ku, Yokohama, Kanagawa 221-8686}

\author[0000-0002-3531-7863]{Yoshiki Toba}
\affiliation{Department of Physical Sciences, Ritsumeikan University, 1-1-1 Noji-higashi, Kusatsu, Shiga 525-8577, Japan}
\affiliation{Academia Sinica Institute of Astronomy and Astrophysics, 11F of Astronomy-Mathematics Building, AS/NTU, No.1, Section 4, Roosevelt Road, Taipei 10617, Taiwan}
\affiliation{Research Center for Space and Cosmic Evolution, Ehime University, 2-5 Bunkyo-cho, Matsuyama, Ehime 790-8577, Japan}


\begin{abstract}
JWST surveys have identified a new class of active galactic nuclei (AGN) called little red dots (LRDs).
Their observational properties challenge the canonical AGN paradigm and provide key insights into the early growth phase of the supermassive black holes (SMBHs).
We report Subaru/PFS spectroscopic follow-up of a radio-loud quasar at $z=1.715$ from the UNVEIL radio AGN catalog and with X-ray detections.
The spectrum displays broad \ciii and \mgtwo emission lines with FWHM $\gtrsim4000~\kms$, accompanied by narrow absorption features.
The spectrum reveals a characteristic $\Lambda$-shape over the rest-frame wavelength ranging $\sim1500$--3500~\AA.
The underlying continuum cannot be reproduced by simply applying dust extinction to typical unobscured quasars. 
Alternatively, it is well described by a blackbody spectrum with a temperature of $T\sim10^4$~K.
This result agrees well with its UV to MIR photometry, which can be well modeled by three blackbody components representing the BH envelope ($T\approx9700$~K), dust torus ($T\approx1500$~K), and host galaxy dust ($T\sim80$~K).
The source is marginally detected in the GALEX NUV, revealing a potential V-shaped spectral energy distribution around 1400~\AA, reminiscent of the spectral feature reported for LRDs whose V-shapes occur around 3000--4000~\AA.
This wavelength shift is broadly consistent with the temperature contrast between our blackbody component ($T\sim10^4$~K) and the lower effective temperature of $T\sim5000$~K expected for the BH envelope of LRDs.
These properties suggest that this source might be caught in an evolutionary phase in which the dense gas envelope characteristic of LRD has begun to fragment, allowing us to witness the emergence of a quasar from an LRD-like state.
\end{abstract}

\keywords{galaxies: active --- 
galaxies: nuclei ---
quasars: supermassive black holes}



% Section 1
\section{Introduction}\label{sec:intro}

One of the fundamental questions on supermassive black holes (SMBHs)
is how they have grown in mass
from the formation of the first black holes (BHs) in the early Universe into the 
current values \citep[e.g.,][]{ina20}.
Recently, the James Webb Space Telescope (JWST) has
opened a new window into high-redshift extragalactic astronomy,
and one of its most important discoveries is a new population
of active galactic nuclei (AGNs) known as little red dots \citep[LRDs;][]{kov23,mat24}.
LRDs are ubiquitously found at $z>4$ and they are characterized by
the ``V''-shaped rest-UV to optical spectral energy distribution (SED).
They exhibit compact morphologies with red optical continua and broad Balmer emission lines \citep[$>80$\%;][]{koc25,hvi25}, frequently accompanied by narrow absorption features \citep[$>30$\%;][]{juo24,lin24}.
However, most LRDs differ from typical type-1 AGNs in several respects, including the scarcity of UV/optical variability \citep{kok24a,zha25a,zha25b}, the absence of strong X-ray emission \citep{ana24,yue24}, and the weakness or even deficit of mid-infrared (MIR) emission from AGN-heated dust \citep[e.g.,][]{wil24,set25}.
Considering the overall energy budget, these findings suggest that the red optical SED is likely intrinsic rather than primarily driven by dust reddening \citep{li25,che25}.

%----------------------------------------fig:PFSspec+SED-----------------------------------%
\begin{figure*}
\begin{center}
\includegraphics[width=0.98\textwidth]{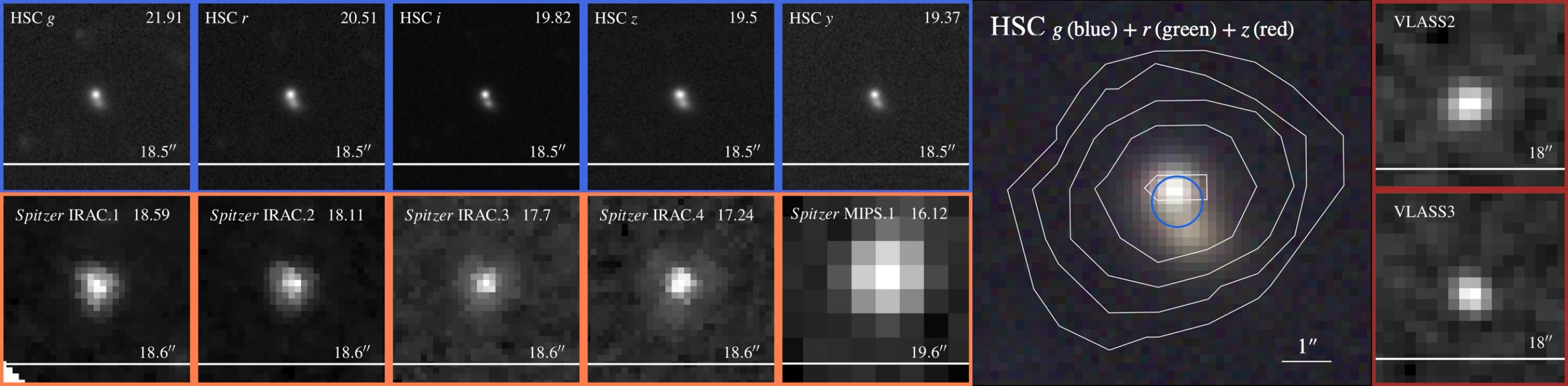}
\includegraphics[width=0.98\textwidth]{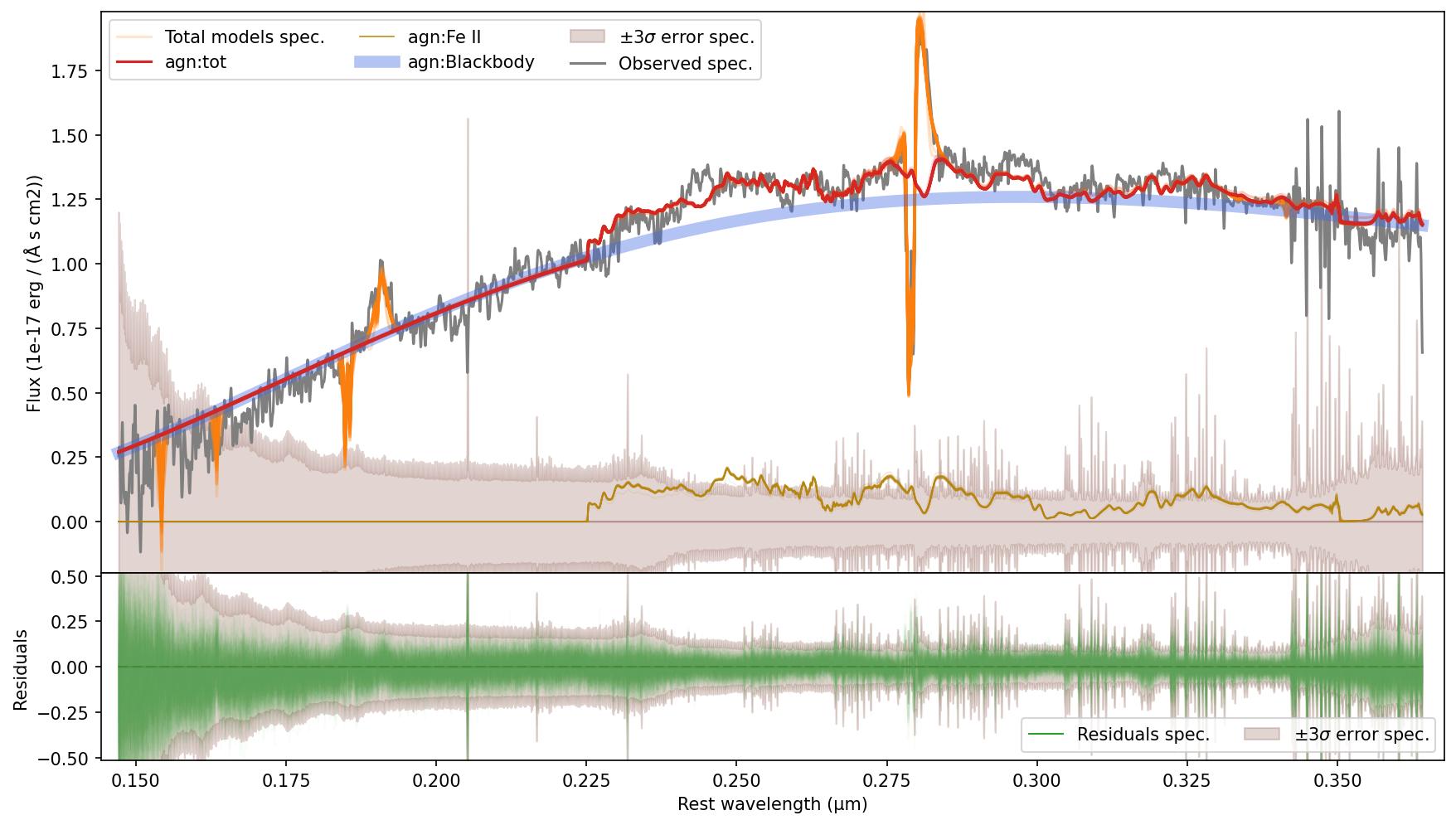}
\includegraphics[width=0.95\textwidth]{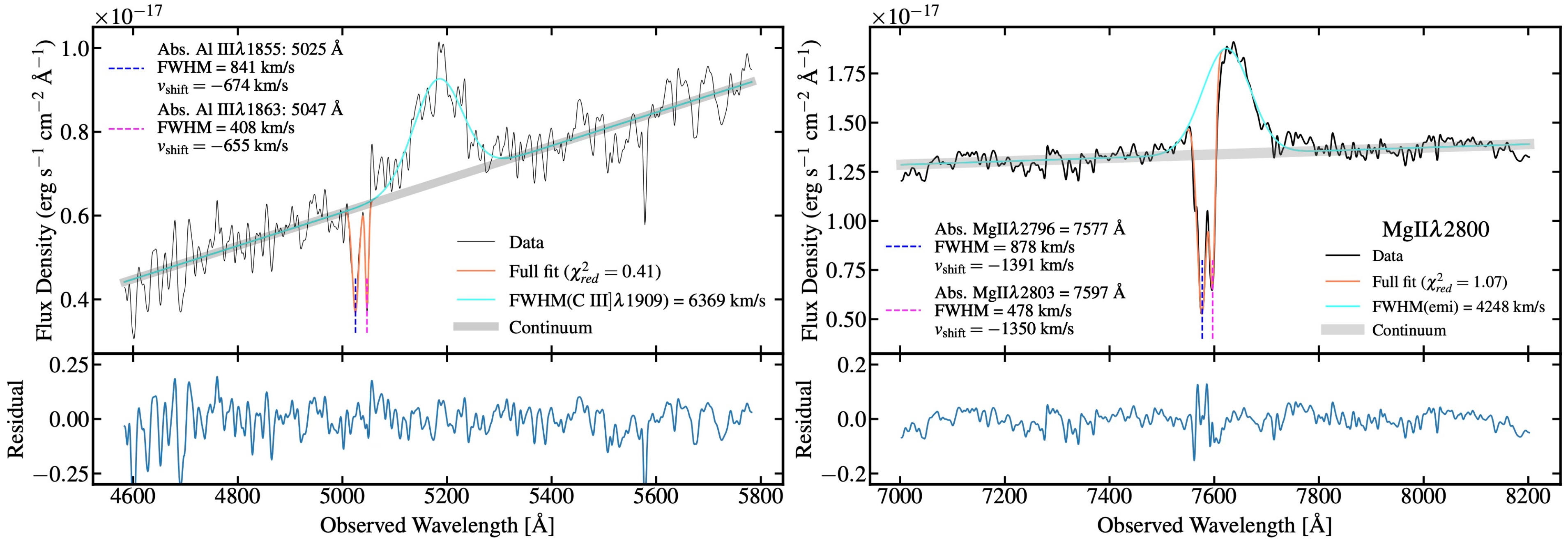}
\caption{
Overview of observations for the radio-loud blackbody quasar \target.
\textit{Top}: Optical images collected from Subaru HSC-SSP, IR images collected from \textit{Spitzer}, and VLASS radio images at 3~GHz observed at Epoch 2 and 3.
The false-color image is rendered using HSC $g+r+z$, overlaid with the white contours representing the VLASS Epoch 2 imaging.
The blue circle represents the PFS fiber with a diameter of $1\farcs02$.
\textit{Middle}: Subaru/PFS spectrum of \target and fitted using $\sanfit$.
\textit{Bottom}: Zoom-in of the Subaru/PFS spectrum \ciii and \mgtwo, as well as best-fits to the underlying continuum and emission/absorption, without incorporating the \feii template.
The spectra are Gaussian-smoothed by every 9 data points.
For absorptions of \aliiidou and \mgtwodou, we show their centers, FWHMs, and velocity shifts in the corresponding panels.
}\label{fig:PFSspec}
\end{center}
\end{figure*}
%----------------------------------------fig:PFSspec+SED-----------------------------------%

%LRDs show black body temperature and would host cocoon
A recent compilation of studies supports a scenario in which the central SMBHs of LRDs are enshrouded by Compton-thick dense gas with column densities of $\NH \gg 10^{24}$~cm$^{-2}$ \citep{kid25,liu2025_lrd_balmer,beg26,nandal26,wendy26,Santarelli26}.
Such a configuration naturally gives rise to a “stellar-like” blackbody spectrum with temperatures of $T \sim 5000$–$7000$~K. 
This temperature range readily accounts for the observed red rest-frame optical continua, while the optically thick gas can produce a Balmer break through absorption by collisionally excited hydrogen atoms populating the $n=2$ level \citep{ina25b, ji25}.
Such an envelope may arise from a direct-collapse BH retaining a remnant envelope \citep{kid25} or in a quasi-star environment for BHs forming through core collapse \citep{beg26}.
These gas-rich configurations can naturally suppress X-ray emission and potentially even radio emission, although super-Eddington accretion has also been proposed as a mechanism to account for the observed X-ray weakness \citep{ina25a}.

% Number density of LRDs drastically decreases at z<3
Another notable property of LRDs is their pronounced cosmic evolution in number density. 
Although LRDs are ubiquitously abundant at $z > 5$, reaching $\sim10^{-4}$~Mpc$^{-3}$ with a possible peak around $z \sim 7$ \citep{tan25}, recent JWST and wide-area ground-based surveys, such as Subaru/HSC, indicate a sharp decline in their number density at $z < 4$ \citep{ma25a}.
\citet{ina25c} proposed that LRDs represent a stochastic phase associated with the early growth of BHs, thus naturally reproducing the observed decline in their number density at $z < 4$. 
Interestingly, \cite{ma25b} also discovered a sharp luminosity cutoff in the LRD luminosity function at $L_\mathrm{bol} \approx 10^{45}$~erg~s$^{-1}$. 
Assuming Eddington-limited accretion, this luminosity corresponds to $M_\mathrm{BH} \lesssim 10^{7}~M_{\odot}$, consistent with expectations from the gas-envelope model.
This mass limit is also consistent with the maximum black hole mass expected from growth over a Salpeter timescale, which is comparable to the inferred LRD lifetime of $\sim 3 \times 10^{7}$~yr \citep{kid25}.
This agreement suggests that the LRD phase may correspond to the initial episode of BH growth in the early Universe, particularly at $z > 4$.

%Cocoon and transient phase should be studied
In this framework, LRDs are characterized by gas-enshrouded accretion with a lifetime of $\sim 10^{7}$~yr. 
After a such enshrouded phase, the envelope should begin to disperse once the SMBH accretion rate exceeds the gas inflow rate from the host galaxy \citep{kid25}. 
In this stage, inflow from the interstellar medium (ISM) is hindered by radiation from the envelope, and the photospheric temperature rises to $\sim 10^{4}$~K \citep{hos12,hos13}.
This hotter-envelope phase may represent the final stage of the LRD episode, preceding the transition to an unobscured quasar. 
As the gas envelope disperses, the central engine becomes progressively exposed, allowing previously suppressed X-ray and possibly radio emission to emerge.
Recently, two studies have reported LRDs, with detections of X-ray and/or radio emissions, in a possible transient phase, while both studies still have an expected envelope temperature of $T=5000$\text{--}$7000$~K \citep{fu25,hvi25}.

In this Letter, we report a serendipitous discovery of Blackbody Quasar and Radio Source (hereafter BBQSORS), a radio and X-ray luminous AGN with merger feature at a spectroscopic redshift of $\tnr{z}{CIII}=1.715$, which is based on the peak wavelength of CIII]$\lambda$1909 emission line. 
\target was originally selected as a radio bright quasar candidate by Very Large Array Sky Survey (VLASS) at 3~GHz \citep{lac20,zhong2025_unveil}.
\target exhibits a blackbody-like UV-to-optical spectrum with $T \sim 10^{4}$~K, which is approximately a factor of two higher than the effective temperatures predicted for the SMBH envelopes proposed for LRDs.
The spectrum is also associated with the broad \mgtwo~emission with a narrow absorber reaching to the continuum level. 
\target also shows a V-shaped SED with a break wavelength around $\lambda\approx1200$~\AA, which is two times shorter than usual turning wavelength ($\lambda\approx3000\text{--}4000$~\AA) of LRD-like V-shape.
We discuss the origin of the spectral features of \target and propose that \target might be in a transient phase of the LRD-like cocoon to an unobscured quasar. 
We adopt the following cosmological parameters throughout this paper: 
$H_0 = 68$\,km\,s$^{-1}$\,Mpc$^{-1}$, $\Omega_\mathrm{M}=0.31$, and $\Omega_\Lambda=0.69$ \citep{pla20}.

% Section 2
\section{Multiwavelength Properties, Subaru/PFS Spectroscopy, and Results}\label{sec:analysis}

\subsection{Morphology}
We show cutouts of \target in key wavelength bands in the top panel of Figure~\ref{fig:PFSspec}.
Subaru/HSC cutouts \citep{hsc_ssp2018} reveal that there are two sources separated by $\sim0\farcs7$ (corresponding to $\sim6$~kpc).
The source in the top left represents \target and it coincides precisely with the centroid of 3~GHz radio emission observed by VLASS in Epochs 2 and 3.
Because of the small separation between the two sources and the fact that the brightness of \target clearly dominates the system, the entire system has been treated as a single quasar, with a centroid offset from \target by approximately $0\farcs2$.

We performed aperture photometry on HSC cutouts for \target and its adjacent source by putting apertures on their centroids with a diameter of $0.5''$ and found that their photometry shows similar trends from $g$ to $z$.
As also supported by the similar colors in the HSC false-color image, this likely rules out the possibility that these two sources are located at very different redshifts.
Therefore, throughout this letter, we treat \target as a radio quasar that resides in a galaxy merger.

\subsection{Subaru/PFS Spectroscopy and Spectral Fitting}

The Subaru/\={O}nohi`ula Prime Focus Spectrograph (hereafter, PFS) UV-optical \citep{tam24} Spectroscopy for \target was obtained as a part of the community filler program (S25A0043QF PI: T. Nagao), which aims to re-observe already known quasars, of the Subaru/PFS Open Use program (Tanaka et al. in prep.).
\target was included in the sample due to its classification as a quasar in Sloan Digital Sky Survey \citep{schneider2010_sdss}, and it was further identified to be radio-loud ($\tnr{L}{3GHz}>10^{26}~\whz$) in the UNIONS-VLASS radio AGN catalog \citep{zhong2025_unveil,gwyn2025_unions}.
The centroid of the PFS fiber is based on the centroid of SDSS imaging and thus has an offset of $\sim0\farcs2$ from \target and of $\sim0\farcs6$ from the companion galaxy.
Since PFS has a core fiber diameter of $1\farcs02$ \citep{tam24} and the typical seeing of PFS observations is $<1$~arcsec, significant contamination of the nearby source is unlikely (see \S\ref{sec: photo} for more information).

% The object ID of \target is \texttt{80141911595205452}.
% The total exposure time was XXX~sec\todo.
The PFS spectrum covers the observed wavelength
of $3800~\AA < \lambda < 12600~\AA$ with three separate spectrograph
modules in the blue (3800--6500~\AA), red (6300--9700~\AA), and near-infrared (NIR; 9400-12600~\AA) arms simultaneously with a spectral resolution of $R\sim3000$.
Data reduction was followed by a standard procedure of the PFS data reduction pipeline (M.~Tanaka et al. in prep.).
Currently, the PFS pipeline suffers from issues in flux calibration and sky subtraction in the NIR arm, resulting in negative flux values. 
We therefore restrict our analysis to wavelengths below $9900$~\AA.

The PFS spectrum (solid gray line) is shown in the middle panel of Figure~\ref{fig:PFSspec} along with its best-fits using $\sanfit$ \citep{s3fit}, and the fitting results are summarized in Table~\ref{tab:properties}.
This spectrum exhibits three distinct features.
First, the spectrum shows broad emission lines in \ciii and \mgtwo with $\mathrm{FWHM>4000}~\kms$ (bottom panel of Figure~\ref{fig:PFSspec}), without considering the strong \feii emission around \mgtwo.
On the blue side of \ciii, absorption features associated with the \aliiidou doublet are detected, with velocity shifts of $v_\mathrm{shift} = -674\pm298\ \kms$ and $v_\mathrm{shift} = -655\pm119\ \kms$, respectively, and a rest-frame separation of 8.1~\AA.
There also exist narrow absorbers associated with the MgII emission line with centers of 7576~\AA\ and 7597~\AA, respectively, in the observed-frame.
The rest-frame separation between these two absorptions is about 7.4~\AA\ adopting $z=1.715$, consistent with that of the \mgtwodou doublet.
The \mgtwodou absorptions have velocity shifts of $v_\mathrm{shift}=-1391\pm118\ \kms$ and $v_\mathrm{shift}=-1350\pm35\ \kms$ with respect to the doublets, respectively, in accordance with the outflow scenario.
These narrow absorbers reach below the continuum level, indicating that the underlying continuum is unlikely to originate from host galaxy stellar emission.

The second feature is a clear $\Lambda$-shaped continuum, which is also present in the SDSS and DESI spectra of \target, indicating that it is robust and independent of the data reduction.
If the UV continuum is interpreted as that of a typical unobscured quasar suffering dust extinction, modeled with a single power-law (PL) representing the accretion disk and Balmer continuum, the best-fit $\sanfit$ model suggests a significant reddening $\av=1.63$.
However, in this scenario, the dust-extincted UV emission from the accretion disk contributes only about 20\% to the observed PFS spectrum, while the UV flux density is dominated by the Balmer continuum.
These results indicate that such a model is physically unrealistic.
Therefore, this strongly curved spectrum is hard to reproduce the known blue spectral shape of unobscured quasars by simply considering a de-reddening by dust extinction, which will be further discussed in \S\ref{sec: rqso}.
Alternatively, given that the Balmer continuum is often approximated in a blackbody form, the spectrum can be best described by a simple blackbody spectrum with a temperature of $T=10471\pm493$~K with a moderate dust extinction of $\av=0.30\pm0.22$, as highlighted by the blue shaded region in the middle panel of Figure~\ref{fig:PFSspec}, which agrees well with the observed photometry in the wider wavelength range (see \S\ref{sec: photo}).

Strong \feii continuum emission (shown by the golden solid lines in the middle panel of Figure~\ref{fig:PFSspec}), particularly around $\sim2800$~\AA\ and on the blue side of \mgtwo, constitutes the third prominent spectral feature.
It is fitted adopting the \feii template of the narrow line Seyfert-I galaxy, I~Zw~1, which are combined from the works of \citet{vest2001_feii}, \citet{veron2004_feii}, and \citet{tsuzuki2006_feii}; these templates are all convolved to FWHM of $1100\ \kms$, i.e., the intrinsic FWHM of the employed combined template.
Such a powerful \feii emission may originate from collisional excitation attributed to low-ionization quasar outflows \citep{wang16_fe_bal}, supported by the \mgtwodou absorption doublet.
This suggests that, for \target, it has abundant \feii in the expected BLR, indicating a relatively metal-rich environment.
% Recently, thanks to higher resolution spectroscopy by JWST/NIRSpec, 
% several LRDs (A2744-45924, \citealt{lab24}; CAPERS-LRD-z9, \citealt{tay25}) are now known to show \feii lines in UV range \citep[notably,
% 2000--3600\AA;][]{lab24,tay25}, but optical \feii lines are weak, mainly due to the low metallicity in the BLR \citep{kov23,tre25}.

%--------------------------------table:photo---------------------------------%
\begin{table}[tp!]
\caption{Summary of the fitting results and physical properties}
\begin{center}
\begin{tabularx}{\columnwidth}{l@{\hspace{2.2cm}}l}
% \begin{tabularx}{\textwidth}{lX}
\hline
\multicolumn{2}{c}{\target, $\tnr{z}{CIII}=1.715^{(a)}$}\\
\multicolumn{2}{c}{$\mathrm{RA}=214.79836$, $\mathrm{Dec}=52.09600^{(b)}$}\\
\hline
\multicolumn{2}{c}{PFS}\\
$\tnr{T}{env,PFS}$ & $10471\pm493$~K \\
$\tnr{L}{env,PFS}$ &  $(4.2\pm1.0)\times10^{45}\ \ergs$ \\
$\av$ &  $0.30\pm0.22$ \\
FWHM (BLR) & $3685\pm247\ \kms$ \\
$F$(\ciii) & $(17.9\pm2.0)\times10^{-17}\ \mathrm{erg/s/cm^2}$ \\
$F$(\mgtwo) & $(61.9\pm23.3)\times10^{-17}\ \mathrm{erg/s/cm^2}$  \\
\noalign{\vskip 2pt}
\hline
$\tnr{T}{env,ph}$ &  $9711^{+300}_{-316}$~K \\
$\tnr{L}{env,ph}$ &  $(9.2\pm0.8)\times10^{45}\ \ergs$ \\
$\tnr{T}{torus}$ &  $1520^{+119}_{-131}$~K \\
$\tnr{L}{torus}$ &  $(5.7\pm0.2)\times10^{45}\ \ergs$ \\
$\tnr{T}{dust}$ &  $81^{+35}_{-13}$~K \\
$\tnr{L}{X}$ &  $2.7^{+1.4}_{-0.8}\times10^{44}\ \ergs$ \\
$\tnr{M}{BH,MgII}^{(c)}$ &  $3.9^{+10.0}_{-2.8}\times10^8\ \msun$ \\
$\tnr{\lambda}{Edd,MgII}^{(c)}$ &  $0.20^{+0.71}_{-0.06}$ \\
\noalign{\vskip 2pt}
\hline
\end{tabularx}
\end{center}
{\vskip 1pt Notes. $^{(a)}$$z=1.720$ based on the broad \mgtwo.
$^{(b)}$Centroid of \target.
Centroids of PFS and the companion are $\mathrm{RA}=214.798335$, $\mathrm{Dec}=52.095943$, and $\mathrm{RA}=214.79823$, $\mathrm{Dec}=52.09579$, respectively.
$^{(c)}$We use $\mathrm{FWHM(MgII)}=3685\ \kms$ corrected for \feii. 
$\tnr{L}{env,PFS}$ and de-reddened $\tnr{L}{2100,PFS}$ with $\av=0.30$ are adopted as conservative estimates.
}\label{tab:properties}
\end{table}
%--------------------------------table:photo---------------------------------%

%----------------------------------------fig:XMM-----------------------------------%
\begin{figure*}[tp!]
\begin{center}
\includegraphics[width=0.92\textwidth]{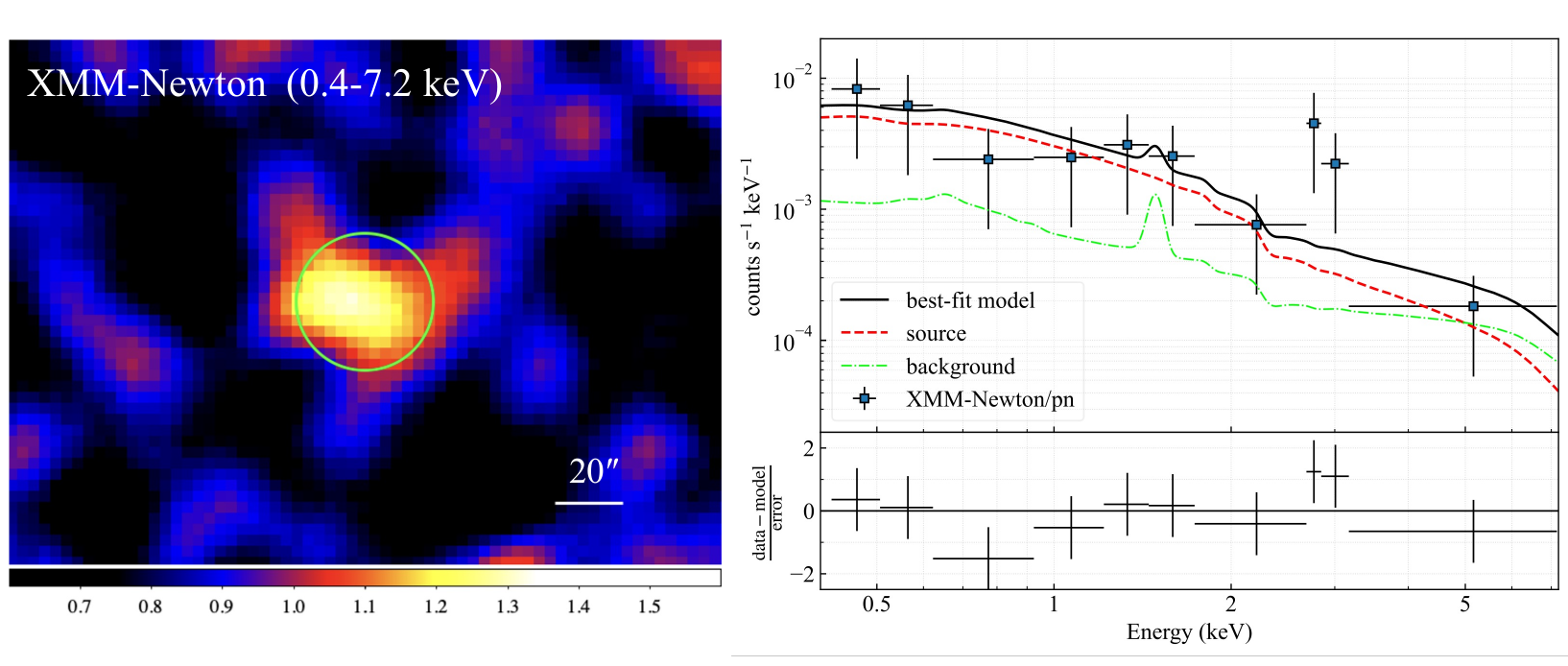}
\caption{
X-ray detection of \target with XMM-Newton. 
The left panel presents the EPIC/pn image in the 0.4–7.2~keV (rest-frame $\sim$1–20~keV) band, detected at a significance of 3.42$\sigma$. 
The green circle indicates the source region with a radius of 20$''$.
The right panel shows the EPIC/pn spectrum folded with the energy response (blue squares) and the best-fitting model (black curve), consisting of source and background components. 
Residuals between the data and the model are shown in the bottom panel.
}
\label{fig:XMM}
\end{center}
\end{figure*}
%----------------------------------------fig:XMM-----------------------------------%

%Emission lines

% See in PFS spectrum
% [CIII]1909+MgII
% FeII?

% Seen in SPHERE-X spectrum
% Ha6563
% De Graaff et al. (2025), \cite{deg25}

% OI8446: Original AGN studies: \cite{rod02}, \cite{mat08}
% 60% of LRD show OI8446; \cite{tri25},  \cite{deg25}; an indicator of 
% Ionisation state of O i is coupled to H ii via charge-exchange reactions
% and O i emission is therefore expected to correlate with the hydrogen
% recombination lines

%--------------------------------table:photo---------------------------------%
\begin{table}[tp!]
\caption{Summary of the photometries}
\begin{center}
\begin{tabularx}{\columnwidth}{llll}
\hline
Band & Flux$^{(a)}$ [$\mu$Jy] & Error [$\mu$Jy] & Ref. \\
\hline
GALEX NUV & 0.428 & 0.043  & 1 \\
CFHT/MegaCam $u$ & 1.361 & 0.147  & 2 \\
CFHT/MegaCam $r$ & 23.828 & 0.116  & 2 \\
SDSS $g$ & 4.835 & 0.481  & 3 \\
SDSS $r$ & 20.324 & 0.880 & 3 \\
SDSS $i$ & 41.153 & 1.365 & 3 \\
SDSS $z$ & 57.863 & 5.329 & 3 \\
PAN-STARRS \textit{g} & 7.688 & 0.805 & 4 \\
PAN-STARRS \textit{r} & 18.885 & 0.704 & 4 \\
PAN-STARRS \textit{i} & 37.225 & 0.246 & 2 \\
PAN-STARRS \textit{z} & 41.679 & 1.839 & 4 \\
PAN-STARRS \textit{y} & 52.048 & 3.054 & 4 \\
HSC $g$ & 6.243 & 0.600 & this work \\
HSC $r$ & 22.789 & 1.146 & this work \\
HSC $i$ & 42.672 & 1.568 & this work \\
HSC $z$ & 57.450 & 1.822 & this work \\
HSC $y$ & 64.823 & 1.952 & this work \\
WFCAM $J$ & 92.897 & 11.979 & 5 \\
WFCAM $H$ & 90.365 & 18.310 & 5 \\
WFCAM $K$ & 100.925 & 16.732 & 5 \\
WISE 3.4 $\micron$ & 155.946 & 1.905 & 6 \\
WISE 4.6 $\micron$ & 211.453 & 3.956 & 6 \\
WISE 12 $\micron$ & 436.418 & 84.813 & 6 \\
\textit{Spitzer} IRAC 3.6 $\mathrm{\mu m}$ & 132.94 & 0.204 & 7 \\
\textit{Spitzer} IRAC 4.5 $\mathrm{\mu m}$ & 207.16 & 0.335 & 7 \\
\textit{Spitzer} IRAC 5.8 $\mathrm{\mu m}$ & 302.89 & 1.552 & 7 \\
\textit{Spitzer} IRAC 8.0 $\mathrm{\mu m}$ & 462.83 & 2.927 & 7 \\
\textit{Spitzer} MIPS 24 $\mathrm{\mu m}$ & $1.298\times10^{3}$ & $0.130\times10^{3}$ & 7 \\
VLA 3~GHz & $7.005\times10^{3}$ & $0.295\times10^{3}$ & 8 \\
VLA 1.4~GHz & $1.452\times10^{4}$ & $1.452\times10^{3}$ & 8 \\
\hline
\multirow{2}{*}{XMM-Newton} & \multirow{2}{*}{$2.16\times 10^{-14,(b)}$} & \multirow{2}{*}{$0.69\times 10^{-14,(b)}$} & \multirow{2}{*}{this work} \\
0.5--7~keV & & & \\
%\hline
%XMM-Newton 0.5--7~keV} & $2.16\times 10^{-14,(b)}$} &
%$0.69\times 10^{-14,(b)}$} & 
%this work\\
% WENSS 327~MHz & $3.1\times10^{5}$ & $3.1\times10^{4}$ & 9 \\
\hline
\end{tabularx}
\end{center}
{$^{(a)}$ The UV-to-IR photometric data are calculated for \target plus its adjacent source.
$^{(b)}$ The unit of flux and error is erg~s$^{-1}$~cm$^{-2}$}
\\ References: (1) \citet{berk2020_galex}; 
(2) \citet{gwyn2025_unions}; 
(3) \citet{sdss_dr16};
(4) \citet{panstarrs};
(5) \citet{uhs_dr3};
(6) \citet{allwise};
(7) \citet{spitzer_seip};
(8) \citet{zhong2025_unveil}.
% (9) \citet{wenss}.
\label{tab:photometry}
\end{table}
%--------------------------------table:photo---------------------------------%

\subsection{XMM-Newton X-ray Analysis}\label{sec:X-ray}
The target was observed with XMM-Newton, which carries three European Photon Imaging Cameras (EPICs), 
on 2016 January 23 (ObsID 0765080801) and 27 (ObsID 0765080901) for exposures of 20~ks and 14~ks, respectively.  
We processed the data from the two EPIC/MOS cameras and the EPIC/pn camera using the XMM-Newton Science Analysis System (SAS; \citealt{Gab04}) v21.0.0.  
The raw data were processed with the EMPROC and EPPROC scripts, and good events were selected with PATTERN $\leq 12$ for MOS and PATTERN $\leq 4$ for pn.  
Events during background flares were removed using thresholds of 0.15 counts s$^{-1}$ in the ${>}10$~keV band for MOS and 0.3 counts s$^{-1}$ in the 10--12~keV band for pn.  
Source spectra were extracted from a circular region with a radius of $20''$, and background spectra from a nearby region with a radius of $40''$.  
Ancillary response files and response matrix files were generated with ARFGEN and RMFGEN, respectively.  
However, only the pn data from January 27 were usable, as the other observations were affected by high background flares and did not show significant source detection.  
In the left panel of Figure~\ref{fig:XMM}, we present a 0.4--7.2~keV (rest-frame ${\sim}1$--20~keV) pn image, where the source is detected at a significance of $3.42\sigma$.  
The 4XMM catalog (DR9--13; \citealt{Web20}) reports a consistent ${\approx}3\sigma$ detection in the 0.5--4.5~keV band, with no nearby X-ray sources within a $120''$ region.

We simultaneously fitted the source and background spectra using the $C$-statistic.  
Galactic absorption was fixed at a hydrogen column density of $N_{\rm H} = 1.03 \times 10^{20}$~cm$^{-2}$ \citep{Wil13}.  
The source emission was modeled as an absorbed PL ({\tt zTBabs*zpo} in the XSPEC terminology) with the photon index fixed at $\Gamma = 1.8$.  
The background model included Galactic halo emission with a temperature of $\sim$0.25~keV and 0.2 solar metallicity, the cosmic X-ray background (absorbed PL with $\Gamma = 1.46$), the non--X-ray background (PL with $\Gamma = 0.24$), and an instrumental Ar~K$\alpha$ line at 1.487~keV \citep{Lec08,Cova19}.  
Owing to limited photon statistics, only the normalizations of the background components were allowed to vary.

The model adequately reproduces the spectra with 
$\Delta C / \mathrm{d.o.f.} = 34.24 / 35$ (see the right panel of Figure~\ref{fig:XMM}).  
The observed 0.5--7~keV flux is 
$F_{0.5-7~\mathrm{keV}} = (2.16 \pm 0.69) \times 10^{-14}~\mathrm{erg~s^{-1}~cm^{-2}}$.  
The best-fit parameters constrain the column density to 
$N_{\rm H} = 1.7^{+18.0}_{-1.7} \times 10^{21}$~cm$^{-2}$ 
and the intrinsic 2--10 keV luminosity to 
$L_{\rm X} = 2.7^{+1.4}_{-0.8} \times 10^{44}$~erg~s$^{-1}$.  
Adopting $\Gamma = 2.2$ yields a mildly higher obscuration, 
$N_{\rm H} \sim 1 \times 10^{22}$~cm$^{-2}$, 
while the luminosity remains nearly unchanged.  
These results indicate that \target hosts a luminous quasar
with a very mild obscuration of $N_\mathrm{H}\sim10^{21-22}$~cm$^{-2}$.

%----------------------------------------fig:SED-----------------------------------%
\begin{figure*}[tp!]
\begin{center}
\includegraphics[width=0.9\textwidth]{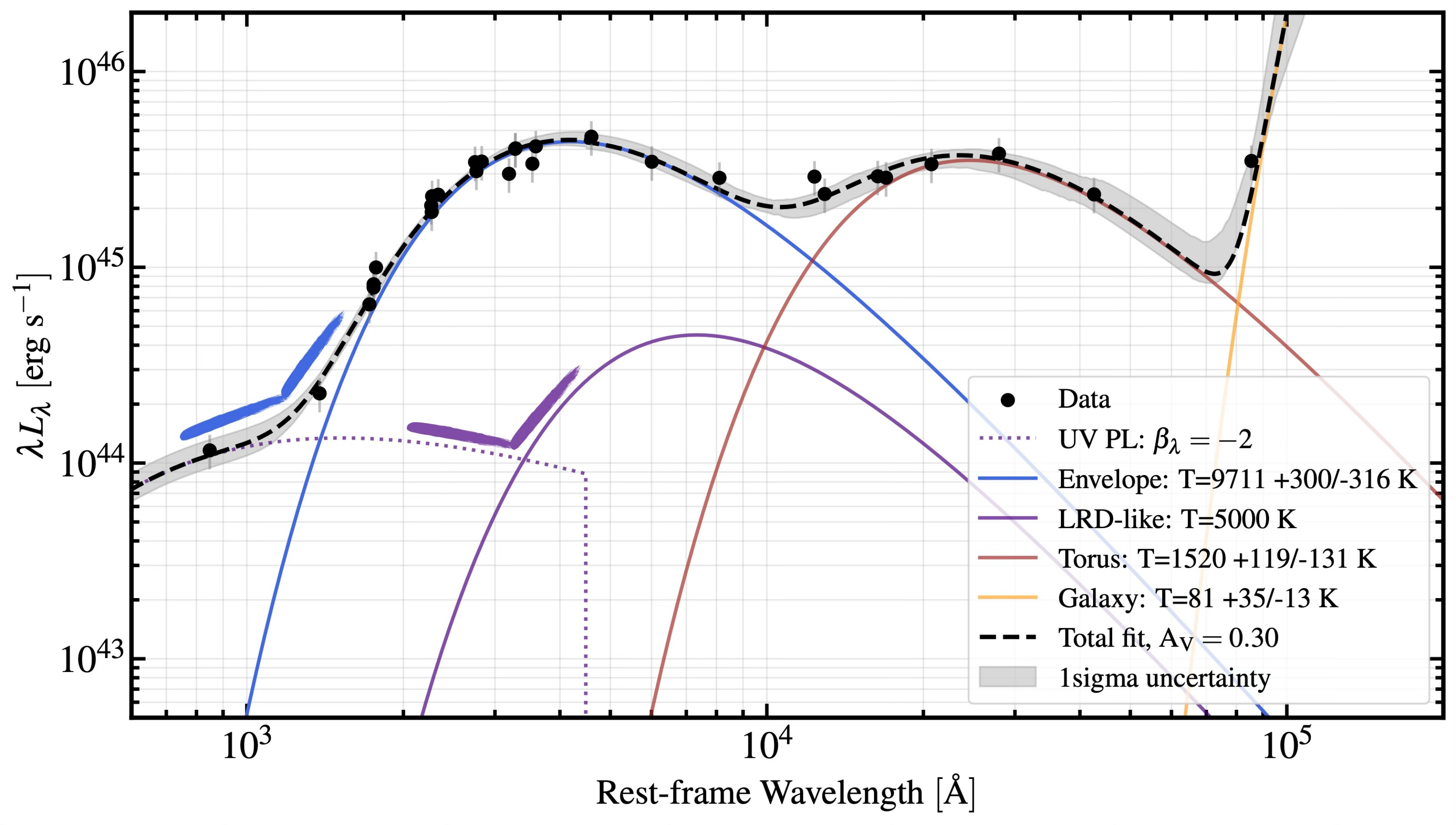}
\caption{
SED fitting for UV-to-MIR photometry with three blackbody components representing the SMBH envelope (blue), AGN dust torus (brown), and galactic dust (golden), plus a UV PL with fixed $\tnr{\beta}{\lambda}-2$, assuming $\av=0.30$ derived from the PFS spectrum (purple dotted line).
The purple solid line represents an LRD-like MBB component with $T = 5000$~K and $\tnr{\beta}{MBB} = 0$, with its amplitude scaled according to $L \propto T^4$.
The dust-extincted UV PL continuum intersects with the blackbody component originating from the accretion disk/LRD envelope, reproducing the LRD-like V-shape around $\lambda\approx3400~\AA$.
}
\label{fig:SED}
\end{center}
\end{figure*}
%----------------------------------------fig:SED-----------------------------------%

\subsection{UV to MIR Photometry}\label{sec: photo}
Multiwavelength photometry has been compiled from UV to MIR (up to 24~$\mu$m), as well as from the radio (1.4~GHz and 3~GHz) and X-ray bands, which are summarized in Table~\ref{tab:photometry}.
Because of the small separation ($0\farcs7$) between \target and its adjacent source, the recorded photometric data treat the two sources as an entity.
Although HSC resolves the system into separate components, we performed forced photometry centered on the PFS fiber position using an aperture diameter of $2\farcs4$ to enable a consistent comparison with the archived data, and the resulting measurements are listed in Table~\ref{tab:photometry}.
Alternatively, placing the aperture on the centroid of \target results in fractional differences of as large as $\sim6\%$ across the $g$ to $y$ bands.
We also performed a test by placing photometric apertures at the centroids of \target\ and the adjacent galaxy, respectively, adopting aperture radii ranging from $0\farcs05$ to $0\farcs4$, given their separation of $\sim0\farcs7$. 
The flux density ratio of \target\ to the galaxy reaches a maximum value of $\sim3$ at $0\farcs1$ and remains above 2 up to $0\farcs4$.
Therefore, significant contamination from the adjacent galaxy is unlikely.
Nevertheless, to conservatively account for any residual contribution from the neighboring source, we adopted a uniform 20\% uncertainty on the observed flux densities in the SED fitting.

Because a high-temperature blackbody subject to stronger dust extinction can produce a spectral shape similar to that of a lower-temperature blackbody with weaker extinction \citep[e.g.,][]{kid25},
we additionally applied a power-law (PL) dust-extinction curve, $A_\lambda \propto \lambda^{-1}$ \citep[e.g.,][]{fawcett2022_rqso}, to the blackbody components.
We adopted $\av = 0.30$, derived from the best-fit model to the PFS spectrum using $\sanfit$, which also assumes a PL extinction law.

Figure~\ref{fig:SED} shows the obtained UV to MIR SED and the best-fit.
Similarly to the PFS spectrum that can be fitted with a single blackbody spectrum of $\tnr{T}{env,PFS}\sim10000$~K, the rest-frame UV--optical photometry ranging 0.1--0.8~\micron\ can also be described by a blackbody component of $\tnr{T}{env,ph}=9711^{+300}_{-316}$~K representing an optically thick gaseous structure surrounding the SMBH, i.e, the SMBH envelope.
The IR continuum up to about 5~\micron\ can be naturally reproduced by involving an AGN dust torus of $\tnr{T}{torus}\approx1500$~K.
The increase in the flux density at $\tnr{\lambda}{obs}=24$~\micron\ can be explained by introducing the galactic dust.
Lacking IR observation at longer wavelengths, the galactic dust temperature is left poorly constrained to be $\tnr{T}{dust}\approx80$~K.

Figure~\ref{fig:SED} also shows that the SED of \target has an excess at $\tnr{\lambda}{rest}\approx850$~\AA.
This flux density is marginally above the detection limit of GALEX NUV \citep[25.1 AB mag;][]{berk2020_galex}.
This excess cannot be explained by a single blackbody component in the UV regime.
We then assumed a UV continuum in the PL form adopting a spectral slope of $\tnr{\beta}{\lambda}=-2$ (defined as $\tnr{F}{\mathrm{\lambda}}\propto\mathrm{\lambda^{\beta_\lambda}}$) that is commonly observed for LRDs \citep{kov23}, which is consistent with the non-detection in the GALEX FUV band \citep[24.0 AB mag;][]{berk2020_galex}.
This UV PL component is incorporated into the SED fitting with $\av=0.30$.
The resulting dust-extincted UV PL intersects with a blackbody component of $\tnr{T}{env,ph}\approx9700$~K, producing an LRD-like V-shaped feature, but around $\tnr{\lambda}{obs}\approx1400$~\AA.

The optical-to-IR continua of LRDs are often modeled using the modified blackbody (MBB) model that depends on a PL slope $\tnr{\beta}{MBB}$ \citep{deg25}, and our model can be translated into a MBB one with $\beta_{\mathrm{MBB}} = 0$.
Assuming an LRD-like blackbody component with $T = 5000$~K and $\beta_{\mathrm{MBB}} = 0$ \citep{deg25}, with the amplitude scaled according to $L \propto T^4$, we find that this model reproduces the characteristic LRD V-shaped feature at a rest-frame wavelength of $\lambda_{\mathrm{rest}} \approx 3400$~\AA\ (purple solid and dotted lines in Figure~\ref{fig:SED}).
Given the marginal detection of NUV and the potential contamination from the companion galaxy, the robustness and origin of this UV excess remain unclear, and future observations are required to constrain the emission origins.

%------------------fig: LRD selection misses BB-----------------%
\begin{figure}
\begin{center}
\includegraphics[width=1\columnwidth]{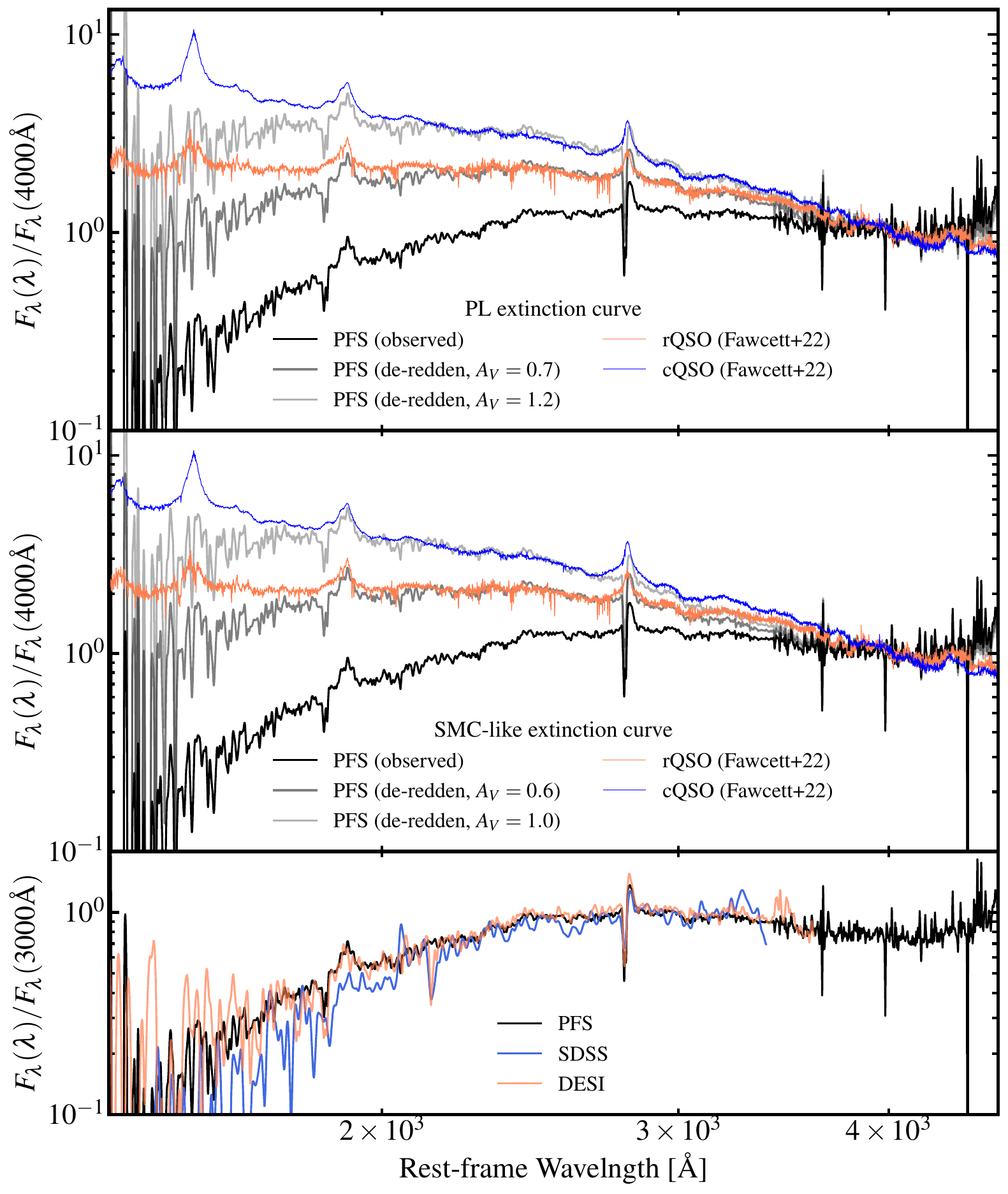}
\caption{
\textit{Top}: Comparison of the spectrum for \target with the median spectra of typical blue quasars (cQSOs) and of red quasars (rQSOs) suffering moderate dust reddening \citep{fawcett2022_rqso}, where all spectra are normalized at 4000~\AA.
We de-redden the PFS spectrum assuming a PL dust extinction curve with $\av=0.7$ and $\av=1.2$.
\textit{Middle}: The same as the top panel, while the de-reddening adopts an SMC-like extinction curve.
Although the de-reddened PFS spectrum have similar spectral shapes with cQSOs and rQSOs from about 2200~\AA\ to 4000~\AA, there are large discrepancies at the blue end for both PL and SMC-like extinction curves.
\textit{Lower}: Comparison of PFS, SDSS, and DESI spectra for \target, where all are normalized at 3000~\AA.
\label{fig:rqso}}
\end{center}
\end{figure}
%------------------fig: LRD selection misses BB-----------------%

\section{Discussion}\label{sec:discussion}

\subsection{Is \target red quasar or a different population?}
\label{sec: rqso}

Based on the large spectroscopic survey of SDSS, a subset of quasars was identified because of their extreme red color in the UV continuum, which exhibits a clear curvature and cannot be well explained by a PL continuum \citep{richards2003_red_qso}.
This population is commonly referred to as red quasars (hereafter rQSOs), the majority of which are reddened by intervening dust \citep{richards2003_red_qso, hopkins2004_rqso,klindt2019_rqso}.
The levels of dust extinction range from up to $\av=0.7$ for those at $z\sim1.5$ \citep{fawcett2022_rqso} to $E(B-V)=1.5$ for the extremely dust-reddened populations \citep{urrutia2009_rqso}.
Additionally, rQSOs have a high incidence in moderate radio luminosity of $\lfirst\sim10^{25\text{--}27}\ \whz$ \citep{klindt2019_rqso,rosario2021_rqso}.

\target has a 1.4~GHz spectral luminosity of $\lfirst\sim2\times10^{26}\ \whz$. 
It exhibits an exceptionally steep UV spectral slope across 1000--2000~\AA\ in both spectroscopic and photometric data ($g-i>2$).
This is clearly demonstrated by the comparison between \target and the median spectrum of normal quasars shown in Figures~\ref{fig:rqso}.
These properties are in line with the characteristics of rQSOs.
We therefore compare the PFS, SDSS, and DESI spectra of \target with the median spectra of typical blue quasars (cQSOs) and of red quasars (rQSOs) that are moderately dust-reddened \citep{fawcett2022_rqso} by normalizing all spectra to 4000~\AA.
We de-redden the PFS spectrum assuming a dust-extinction curve in the PL form.
At $\sim2200$--4000~\AA, the de-reddened PFS spectrum agrees well with rQSOs assuming $\av=0.7$ and shows good consistency with cQSOs assuming $\av=1.2$.
However, at shorter wavelengths ($\tnr{\lambda}{rest}\lesssim2000$~\AA), there exist large discrepancies between the de-reddened PFS spectra and cQSOs and rQSOs.
Adopting an SMC-like dust extinction curve \citep{Gordon2024_dust,gordon24_smc} instead yields similar results while the corresponding $\av$ could be slightly smaller than those of the PL form.
A larger dust extinction can produce the blue end of the \target spectrum with a reddened cQSO, but then introduces significant discrepancies over the $\sim1600$–3000~\AA\ range.

These large discrepancies come from the highly curved spectral feature of \target, which is robust for its existence in PFS, SDSS, and DESI.
In conclusion, the blackbody quasar \target mimics rQSOs rather than being the exact same population, although it may represent the progenitors of rQSOs, depending on the evolution of the structures surrounding the SMBH.

%------------------fig: BBQSORS Evolution-----------------%
\begin{figure}
\begin{center}
\includegraphics[width=1\columnwidth]{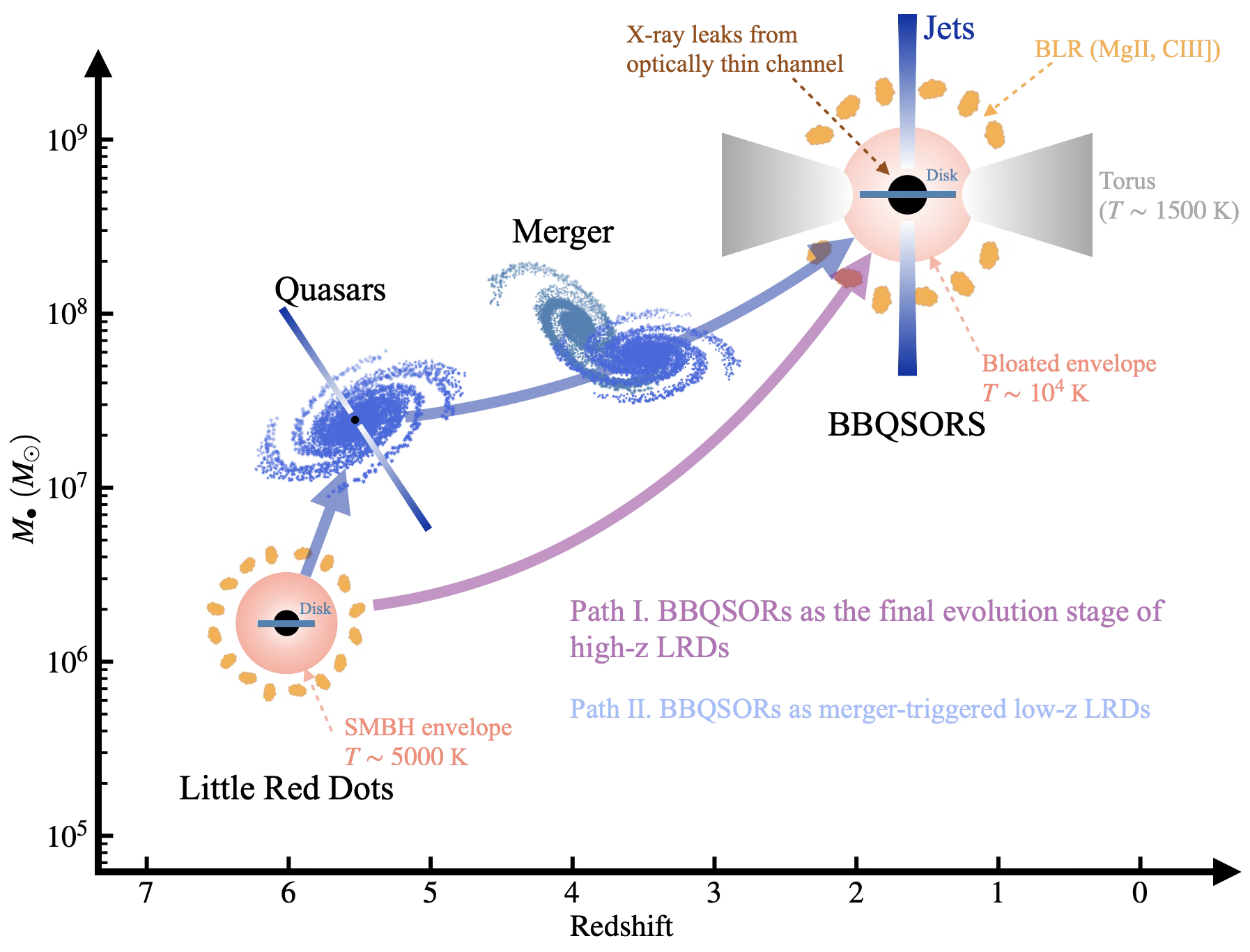}
\caption{
Schematic illustration of two possible evolutionary pathways linking 
LRDs and BBQSORS discussed in Section~3.2.1.
In Path I, BBQSORS represent the final stage of high-$z$ LRD evolution,
in which the cooler SMBH envelope ($T\sim5000$K) evolves into a hotter,
bloated envelope with $T\sim10^4$~K while the system becomes less obscured.
In this stage, broad-line region emission (e.g., Mg\,{\sc ii} and C\,{\sc iii}]) and radio jets become visible, and X-rays can leak through an optically thin channel. 
In Path II, the envelope of BBQSORS arise instead as a result of intense inflow triggered by a merger, and mimicking low-$z$ analogs of LRDs. 
}\label{fig: bbq_evo}
\end{center}
\end{figure}
%------------------fig: BBQSORS Evolution-----------------%

%------------------fig: Photometry comparison with literature-----------------%
\begin{figure*}
\begin{center}
\includegraphics[width=0.98\textwidth]{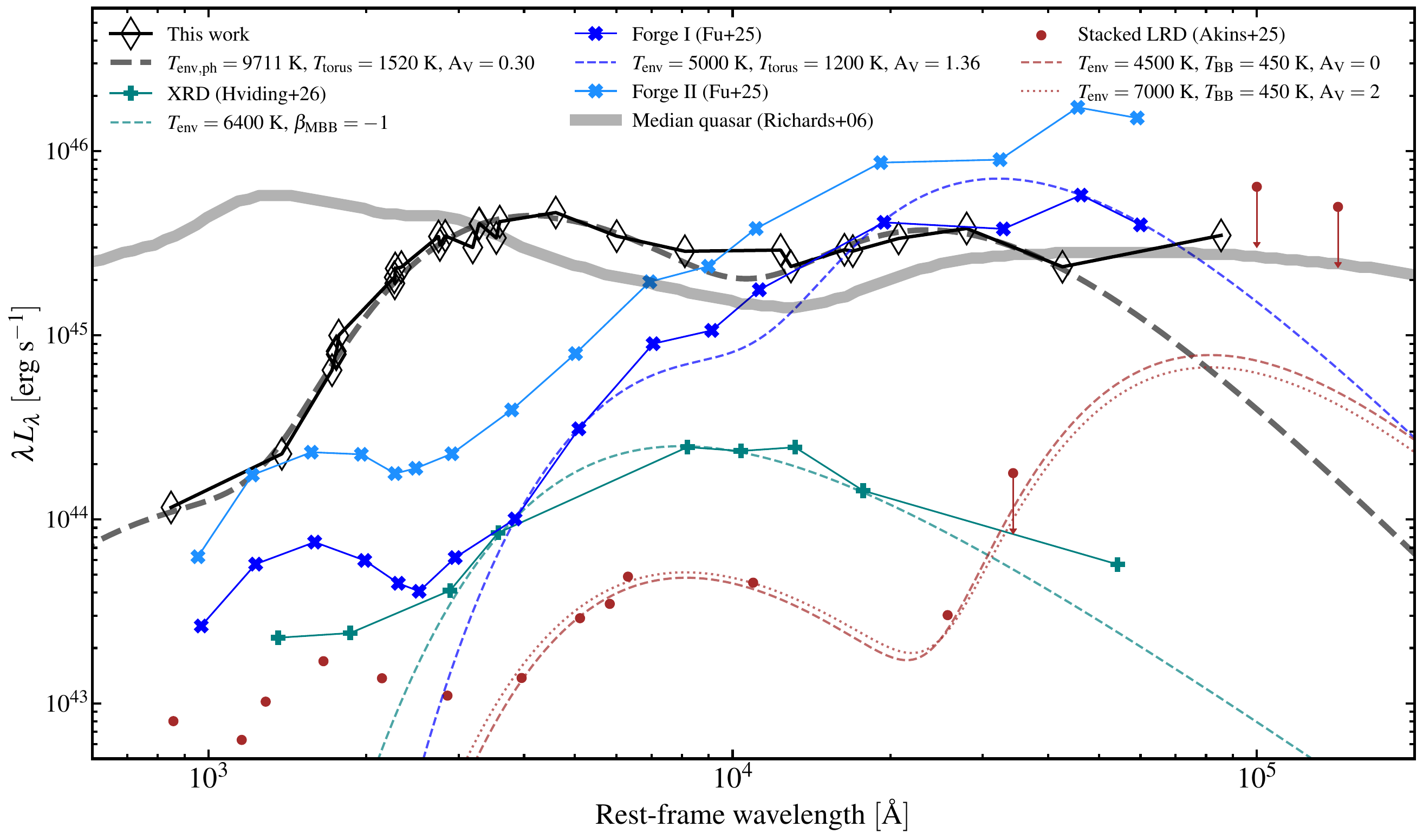}
\caption{
Comparison of the SED of the blackbody quasar \target (this study; black solid line) with those of the X-Ray Dot \citep[moss-green solid line;][]{hvi25}, transitioning LRDs \citep[Forge I and II; blue and cyan lines, respectively;][]{fu25}, and stacked LRDs at $z\sim6$ \citep[brown filled circles;][]{akins25_stack_lrd}.
Their optical-to-IR continua are broadly reproduced by (modified) blackbody models with different temperatures and dust extinctions, shown by the dashed and dotted lines.
For reference, the median spectrum of normal quasars is also overplotted \citep[thick gray solid line;][]{richards06_qso}.
}\label{fig:LRD_SED_comparison}
\end{center}
\end{figure*}
%------------------fig: Photometry comparison with literature-----------------%

\subsection{\target as a transitioning LRD candidate}
\subsubsection{What does $T\approx10^4$~K indicate in the context of LRDs?}

Recent theoretical studies suggest that the red color in the optical of LRDs can be maintained by a massive, optically thick gaseous envelope that surrounds the central accreting black hole and reprocesses the intrinsic emission into a cool optical--NIR SED \citep{ina2025_lrd_review}. 
In particular, \cite{kid25} demonstrated that such an envelope can gravitationally confine radiation-driven outflows launched by super-Eddington accretion, allowing the system to radiate near the Eddington limit while sustaining an extended effective photosphere with temperatures close to the Hayashi limit, $T_{\rm eff}\sim(5$--$7)\times10^3$~K. 
In this regime, the envelope efficiently absorbs and re-emits the central radiation, suppressing strong feedback, and enabling gas to continuously infall from the surrounding ISM.

As the central black hole grows, however, the accretion rate required to sustain near-Eddington luminosities increases in proportion to the black hole mass, whereas the ISM inflow rate feeding the envelope may remain approximately constant.
Once the mass supply becomes insufficient to replenish the envelope losses, the envelope mass gradually declines and the structure becomes patchy, leading to a covering factor below unity \citep{hvi25}.
The X-ray source may therefore become visible through optically thin channels, while radio jets may drill through the envelope and further enhance the detectability of the X-ray emission.
\citet{kid25} further showed that a photosphere can emerge either within the SMBH envelope or in the infalling material. 
When the effective photospheric temperature approaches $T \sim 10^{4}$~K, corresponding to the onset of hydrogen ionization, the envelope structure becomes unstable due to the rapid increase in thermal pressure and enhanced radiation–matter coupling.
At this stage, gravitational confinement weakens, allowing radiation-pressure–driven outflows to escape and ultimately leading to rapid dispersal of the envelope.

This transition near $T\sim10^4$~K therefore marks a critical evolutionary phase, in which the system shifts from an envelope-dominated, reprocessed state to a feedback-dominated phase characterized by strong mass loss and partial exposure of the central engine. 
In this picture, LRDs with cool blackbody-like SEDs trace an early, bloated-envelope phase, while sources exhibiting higher effective temperatures and emerging X-ray or radio emission represent a short-lived blown-out phase preceding the emergence of an unobscured quasar.

We speculate on the evolution path of the blackbody quasar \target in Figure~\ref{fig: bbq_evo}.
In the first scenario, the central SMBHs of LRDs continuously gain their masses, and once the SMBH exceeds a certain threshold such that the SMBH envelope can no longer sustain a steady state, we are witnessing the final evolution stage of LRDs.
In the second scenario, LRDs have already evolved beyond the envelope-enshrouded phase, becoming high-$z$ quasars or AGNs.
Their SMBHs remain at relatively low accretion rates.
Then, at a certain time, they encounter a nearby galaxy and start galaxy merging/interactions.
The intense gas inflow accompanied by these events leads to a massive accretion onto the SMBH.
If the inflowing material exceeds the rate at which the SMBH can consume it, with $\dot{M}\approx 10-100 \msun$~yr$^{-1}$ \citep{kid25}, an optically thick gas envelope can form, pushing the system back into an LRD-like state.
This is close to the maximum theoretically achievable inflow rate \citep{thompson2005,ina16} and
several quasars have been are observationally shown to reach such high accretion rate \citep[e.g.,][]{wu15,obu26}. 
Therefore, this condition may occasionally be realized in extreme merger environments.
This second scenario may also provide an explanation for the LRD-like source at $z\approx0.4$ that also resides in a merger \citep{chen25_lrd}.
%These descriptions remain qualitative, as quantitative analysis is uncertain due to the limitations of the current observations. Follow-up observations of \target, as well as studies of larger samples, will be required to test these scenarios.

%------------------fig: T and L comparisons with LRD-----------------%
\begin{figure*}
\begin{center}
\includegraphics[width=0.98\textwidth]{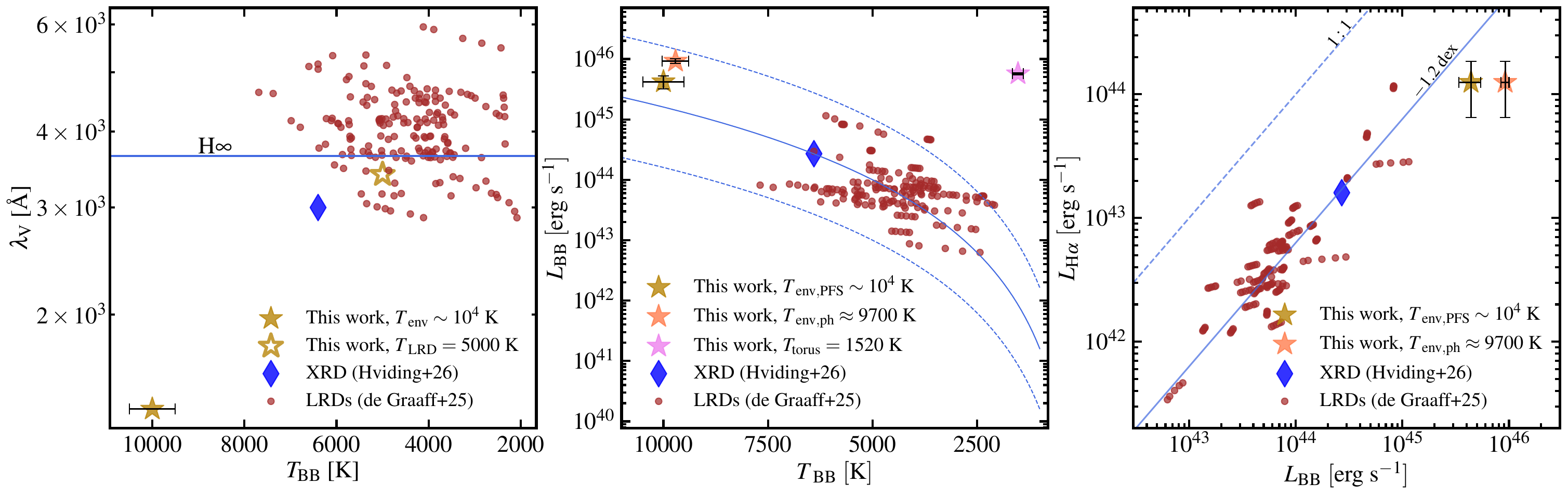}
\caption{\label{fig:LRD_Property_comparison}
Comparison of \target with LRDs at $z<4.5$ from \citet{deg25} and the XRD from \citet{hvi25} in the $\lambda_\mathrm{V}$--$T_\mathrm{BB}$, $L_\mathrm{BB}$--$T_\mathrm{BB}$, and $L_{\halpha}$--$L_\mathrm{BB}$ planes.
%Comparison of the blackbody temperature $\tnr{T}{BB}$ and luminosity $\tnr{L}{BB}$, 
Here, $\tnr{\lambda}{rest}$ represents the wavelength where the V-shaped feature emerges. 
%and the H$\alpha$ luminosity of \target with those of LRDs at $z < 4.5$ from \citet{deg25} and the XRD from \citet{hvi25}.
In the left panel, the solid line indicates the Balmer limit.
In the middle panel, the solid line shows the expected $\tnr{L}{BB}\text{--}\tnr{T}{BB}$ relation for a fiducial LRD-like photospheric radius \citep{ina2025_lrd_review}, with dashed lines indicating offsets of $\pm1$~dex.
In the right panel, the dashed line represents the $1:1$ relation and the solid line shows the same relation shifted by $-1.2$ dex.
}
\end{center}
\end{figure*}
%------------------fig: T and L comparisons with LRD-----------------%

\subsubsection{Comparison with transition-LRD candidates}
A few objects have been argued to be LRDs in transitions, including the X-ray Dot \cite[XRD;]{hvi25} at $z=3.28$ and  two unusual LRDs at $z\approx2.9$ \citep[][hereafter Forge I and II]{fu25}.
We here compare \target as a transition source bridging LRDs and normal quasars with these objects. 
Their photometric data are shown in Figure~\ref{fig:LRD_SED_comparison} by solid lines with markers, overlaid with dashed lines indicating the SED fitting with blackbody components of different effective temperatures and modified by dust extinction.
We also compare these candidates of the transitioning LRD with the stacked LRD photometry at $z=6$ \citep{akins25_stack_lrd}.

In the scenario where LRDs are AGNs whose SMBHs are enshrouded by dense gas envelopes, their stacked photometry can be reproduced by a blackbody component of $\tnr{T}{eff}=4500$~K without visual extinction or of $\tnr{T}{eff}=7000$~K and reddened by $\av=2$ \citep{kid25}.
The absence of an AGN dust torus, or that the dust torus has just started to form, leads to the overall non-detection of radiation from LRDs in the MIR regime.
For XRD and Forge sources, their optical-–NIR continua can be modeled with a single blackbody component at temperatures of $\tnr{T}{eff} \sim (5\text{--}6)\times10^{3}$~K as well.
In the MIR regime, the absence of dedicated MIR observations prevents a firm assessment of whether a dust torus is present in the XRD.
Notably, however, the 1–6~\micron\ photometry of Forge sources exhibits strong similarities to \target\ and can be well reproduced by a blackbody component with $T \sim 1200$~K, suggestive of the emergence of a dust torus (or alternatively $T \sim 500$~K; see \citealt{fu25} for further discussion).
Hence, the XRD and Forges populations are likely to represent an intermediate evolutionary phase as the LRDs transition toward \target, accompanied by the progressive development of a dust torus. %and the expansion of the photosphere.

We estimated the SMBH mass of \target using the broad \mgtwo emission line fitted by $\sanfit$, which has been corrected for the strong \feii emission.
We adopted the empirical relation $\tnr{M}{BH}=10^{6.79\pm0.55}\left[\frac{\mathrm{FWHM(MgII)}}{1000\ \kms}\right]^2\left(\frac{\tnr{L}{2100\AA}}{\ergs}\right)^{0.5}$ \citep{vest09_bhmass}, where $\tnr{L}{2100\AA}$ is calculated from the PFS spectrum considering $\av=0.30$, resulting in $\tnr{M}{BH}=3.9^{+10.0}_{-2.8}\times10^8\ \msun$.
The luminosity of the dust torus is $(5.7\pm0.2)\times10^{45}\ \ergs$.
For the envelope luminosity, we take the best-fit value of the PFS spectrum, which is $\tnr{L}{env,PFS}=(4.2\pm1.0)\times10^{45}\ \ergs$, as a conservative estimate.
Treating the AGN bolometric luminosity as $\tnr{L}{AGN,bol}\sim\tnr{L}{env,PFS}+
\tnr{L}{torus}$, the Eddington ratio of \target is $\tnr{\lambda}{Edd}\approx0.20^{+0.71}_{-0.06}$.
These properties are comparable to those of Forge sources, although it should noted that the mass estimate remains somewhat uncertain because the UV radiation of \target originates from the SMBH envelope rather than from a standard thin disk.
On the other hand, the VLA-COSMOS deep 3~GHz radio observations of the two Forge sources have flux densities of about $\approx15\ \mujy$ and $84\ \mujy$, corresponding to $\tnr{L}{3GHz}\sim1\times10^{24}\ \whz$ and $\tnr{L}{3GHz}\sim6\times10^{24}\ \whz$, respectively, to be compared with $\tnr{L}{3GHz}\sim5\times10^{26}\ \whz$ of \target.
Therefore, the potential merger event of \target may trigger the outburst of the radio jet \citep[e.g.][]{zhong24_dragonfly}, facilitating the fragmentation of the SMBH envelope. 

In Figure~\ref{fig:LRD_Property_comparison}, we further compare the blackbody temperature $\tnr{T}{BB}$ and luminosity $\tnr{L}{BB}$ of the SMBH envelope, the rest-frame wavelength $\tnr{\lambda}{rest}$ of the V-shaped feature, and the $\halpha$ luminosity of \target with corresponding quantities for LRDs at $z < 4.5$ \citep{deg25} and the XRD \citep{hvi25}.
In the middle panel, we consider an LRD-like photosphere with a fiducial radius determined by $\tnr{R}{ph}=\sqrt{\tnr{L}{ph}/4\pi\tnr{\sigma}{SB}T^4_\mathrm{eff}}$ \citep{ina2025_lrd_review}, where $\tnr{\sigma}{SB}$ is the Stefan-Boltzmann constant, assuming $\tnr{L}{ph}=1\times10^{44}\ \ergs$, and $\tnr{T}{eff}=5000$~K.
Fixing $\tnr{R}{ph}$ and assuming $\tnr{L}{ph}\simeq\tnr{L}{BB,LRD}$, we plot the solid line indicating how $\tnr{L}{ph}$ evolves with the effective temperature.
The luminosity of $\halpha$ of \target was estimated from the spectrophotometry of the Spectro-Photometer for the History of the Universe, Epoch of Reionization, and Ices Explorer \citep[SPHEREx;][]{boc26} and fitted by $\sanfit$ (see Appendix~\ref{app: spherex}) without dust-correction.
Owing to the low spectral resolution and noisy flux measurements, the $\halpha$ line has a measured $\mathrm{FWHM} \approx 2100\pm1800\ \kms$. 
This value is in line with those of \ciii and \mgtwo given the $1\sigma$ uncertainty.
Overall, \target lies broadly consistent with the $L_{\mathrm{BB}}$–$T_{\mathrm{BB}}$ and $L_{\halpha}$–$L_{\mathrm{BB}}$ relations established for LRDs, supporting its identification as a strong candidate for an evolved LRD.

%------------------fig: LRD selection misses BB-----------------%
\begin{figure}
\begin{center}
\includegraphics[width=1\columnwidth]{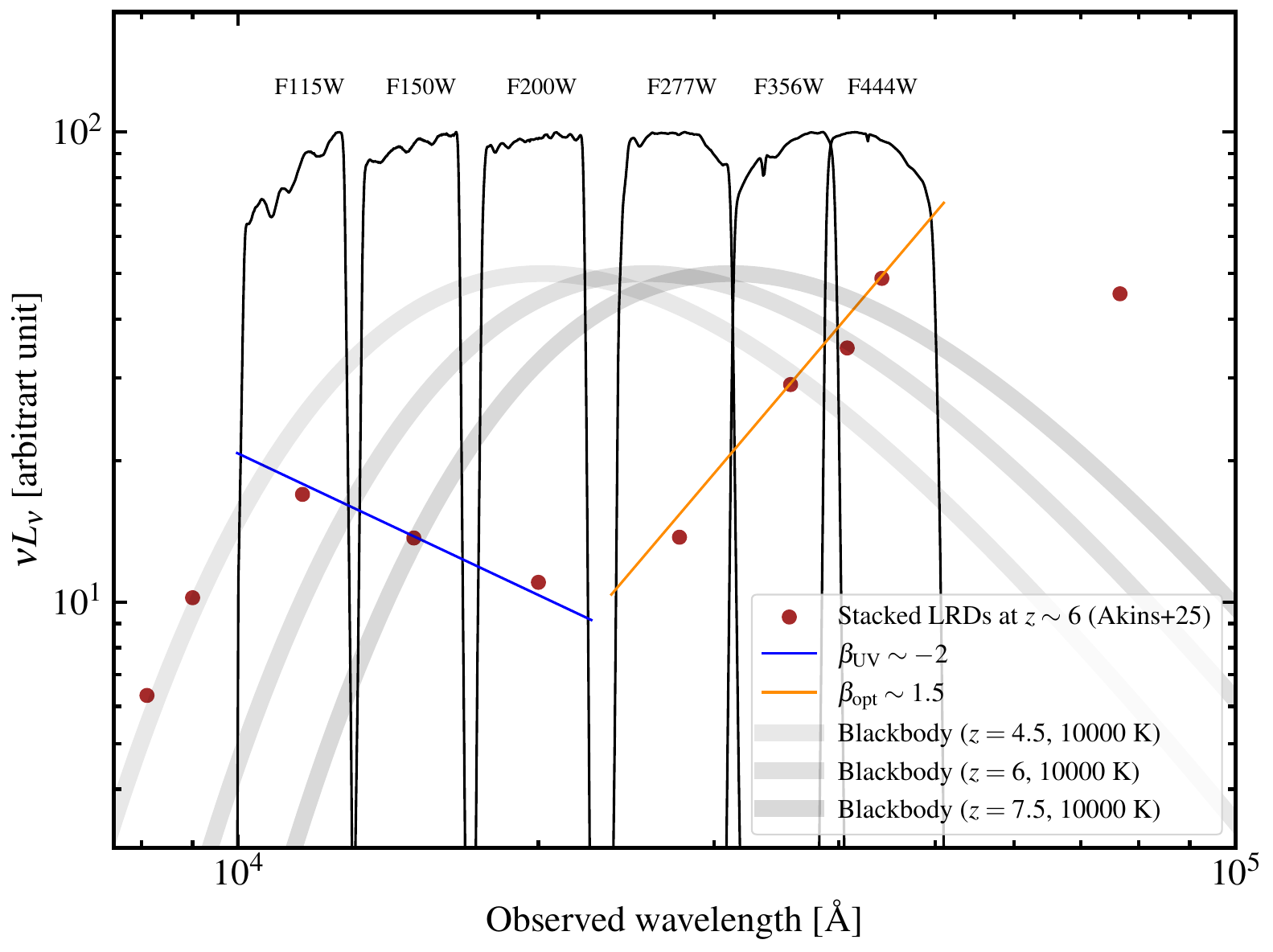}
\caption{
A simple blackbody spectrum with $T=10^4$~K at different redshifts, overlaid on the stacked LRD photometries at $z=6$ \citep{akins25_stack_lrd} and the normalized transmission curves of the JWST NIRCAM filters. While high-$z$ LRDs are selected based on their characteristic V-shaped SEDs  transitioning LRDs characterized by blackbody quasars exhibit the $\Lambda$-shaped SED.
}\label{fig:lrd_miss_bb}
\end{center}
\end{figure}
%------------------fig: LRD selection misses BB-----------------%

\subsection{Why LRD color-color selections do not find $T\sim10^4$K blackbody sources?}

Current LRD color–color selection criteria require four photometric bands sampling the V-shaped feature, two bands on the blue side and additional two bands on the red side, while excluding a band at the V-shaped minimum, in order to mitigate contamination from emission lines \citep[e.g.,][]{kok24,hai25}.
In such a way, the current JWST cannot choose the appropriate $T\sim10^4$ K sources because the bluest JWST/NIRCam band will cover ``bluer'' the $\lya$ break for those at $z\gtrsim6$. 
This makes the color redder in the blue part and bluer in the red part.
The same situation happens if using F115W to F444W to cover the blackbody spectrum, as shown in Figure~\ref{fig:lrd_miss_bb}.
On the contrary of LRDs, this population of possible transitioning LRDs could exhibit a red UV continuum of $\tnr{\beta}{UV}\approx1.5$ and a blue optical continuum of $\tnr{\beta}{opt}\approx-2$.
Resultantly, sources bridging LRDs and normal quasars have been overlooked.

\section{Conclusion}
In this Letter, we report Subaru/PFS observation of a radio-loud quasar, \target, at $\tnr{z}{CIII}=1.715$ and residing in a potential galaxy merger.
The PFS UV spectrum covering $\sim1500$--3500~\AA\ in the rest-frame displays three distinguished features.
The first feature is the presence of broad \ciii and \mgtwo emission lines, with $\mathrm{FWHM} \gtrsim3600\ \kms$, accompanied by \aliiidou and \mgtwodou absorption doublets.
The second feature is the strong \feii continuum emission in the vicinity of \mgtwo.
The third feature is the significantly curved underlying continuum, exhibiting a $\Lambda$-shape.
Globally, the spectrum can be well reproduced by a single blackbody spectrum with an effective temperature of approximately $10000$~K.

The UV--MIR photometric data of \target can be fitted by introducing three blackbody components: an SMBH envelope of $T\approx9700$~K, an AGN dust torus of $T\approx1500$~K, and galactic dust of $T\approx80$~K.
Notably, we found an excess in the flux density of GALEX NUV, which cannot be included in the blackbody model.
By interpolating the GALEX NUV data assuming a power law continuum with $\tnr{\beta}{\lambda} = -2$, the resulting power law spectrum, combining the blackbody spectrum, exhibits the characteristic V-shaped SED observed in LRDs.
Furthermore, when the blackbody spectrum of \target is scaled to an effective temperature representative of LRDs, $T \approx 5000$~K, the UV power law intersects the blackbody component, producing a V-shaped feature around 3400~\AA.

These results suggest that \target is likely to be transitioning from an LRD-like state to a normal quasar.
At this evolutionary stage, the SMBH envelope begins to disperse, as accretion onto the SMBH is no longer sufficient to replenish the envelope mass, and as strong feedback from radiation and radio jets drives further mass loss.
Future NIR observations are needed to search for the H$\alpha$ absorption feature commonly observed in LRDs, as well as the Balmer break expected from an optically thick envelope. 
Measurements of the gas metallicity and direct evidence for outflows would provide further tests of whether \target represents an evolved LRD in a bloated-envelope phase.

\acknowledgments
We thank Jong-Hak Woo for fruitful discussions.
Y.Z. is supported by Japan Society for the Promotion of Science Research Fellowship for Young Scientists.
This work is supported by the Japan Society for the Promotion of Science (JSPS) KAKENHI (25K01043; K.~Ichikawa).
K.I. also acknowledges support from the JST FOREST Program, Grant Number JPMJFR2466 and the Inamori Research Grants, which helped make this research possible.

The Hyper Suprime-Cam (HSC) collaboration includes the astronomical communities of Japan and Taiwan, and Princeton University. The HSC instrumentation and software were developed by the National Astronomical Observatory of Japan (NAOJ), the Kavli Institute for the Physics and Mathematics of the Universe (Kavli IPMU), the University of Tokyo, the High Energy Accelerator Research Organization (KEK), the Academia Sinica Institute for Astronomy and Astrophysics in Taiwan (ASIAA), and Princeton University. Funding was contributed by the FIRST program from the Japanese Cabinet Office, the Ministry of Education, Culture, Sports, Science and Technology (MEXT), the Japan Society for the Promotion of Science (JSPS), Japan Science and Technology Agency (JST), the Toray Science Foundation, NAOJ, Kavli IPMU, KEK, ASIAA, and Princeton University. 

This paper makes use of software developed for Vera C. Rubin Observatory. We thank the Rubin Observatory for making their code available as free software at http://pipelines.lsst.io/.

This paper is based on data collected at the Subaru Telescope and retrieved from the HSC data archive system, which is operated by the Subaru Telescope and Astronomy Data Center (ADC) at NAOJ. Data analysis was in part carried out with the cooperation of Center for Computational Astrophysics (CfCA), NAOJ. We are honored and grateful for the opportunity of observing the Universe from Maunakea, which has the cultural, historical and natural significance in Hawaii.

This work is based on data obtained as part of the Canada-France Imaging Survey, a CFHT large program of the National Research Council of Canada and the French Centre National de la Recherche Scientifique. Based on observations obtained with MegaPrime/MegaCam, a joint project of CFHT and CEA Saclay, at the Canada-France-Hawaii Telescope (CFHT) which is operated by the National Research Council (NRC) of Canada, the Institut National des Science de l’Univers (INSU) of the Centre National de la Recherche Scientifique (CNRS) of France, and the University of Hawaii. 

The Pan-STARRS1 Surveys (PS1) and the PS1 public science archive have been made possible through contributions by the Institute for Astronomy, the University of Hawaii, the Pan-STARRS Project Office, the Max Planck Society and its participating institutes, the Max Planck Institute for Astronomy, Heidelberg, and the Max Planck Institute for Extraterrestrial Physics, Garching, The Johns Hopkins University, Durham University, the University of Edinburgh, the Queen’s University Belfast, the Harvard-Smithsonian Center for Astrophysics, the Las Cumbres Observatory Global Telescope Network Incorporated, the National Central University of Taiwan, the Space Telescope Science Institute, the National Aeronautics and Space Administration under grant No. NNX08AR22G issued through the Planetary Science Division of the NASA Science Mission Directorate, the National Science Foundation grant No. AST-1238877, the University of Maryland, Eotvos Lorand University (ELTE), the Los Alamos National Laboratory, and the Gordon and Betty Moore Foundation.

This publication makes use of data products from the Wide-field Infrared Survey Explorer, which is a joint project of the University of California, Los Angeles, and the Jet Propulsion Laboratory/California Institute of Technology, and NEOWISE, which is a project of the Jet Propulsion Laboratory/California Institute of Technology. WISE and NEOWISE are funded by the National Aeronautics and Space Administration.

This work is based [in part] on observations made with the Spitzer Space Telescope, which was operated by the Jet Propulsion Laboratory, California Institute of Technology under a contract with NASA

This publication makes use of data products from the Spectro-Photometer for the History of the Universe, Epoch of Reionization and Ices Explorer (SPHEREx), which is a joint project of the Jet Propulsion Laboratory and the California Institute of Technology, and is funded by the National Aeronautics and Space Administration.

\facilities{Subaru/PFS, Subaru/HSC, \textit{GALEX}, \textit{CFHT}, \textit{Sloan}, \textit{PS1}, \textit{WISE}, \textit{Spitzer}, \textit{XMM-Newton}, \textit{VLA}}, \textit{SPHEREx}

\appendix
\section{SPHEREx Spectrophotometry}\label{app: spherex}
\begin{figure*}
\begin{center}
\includegraphics[width=0.9\textwidth]{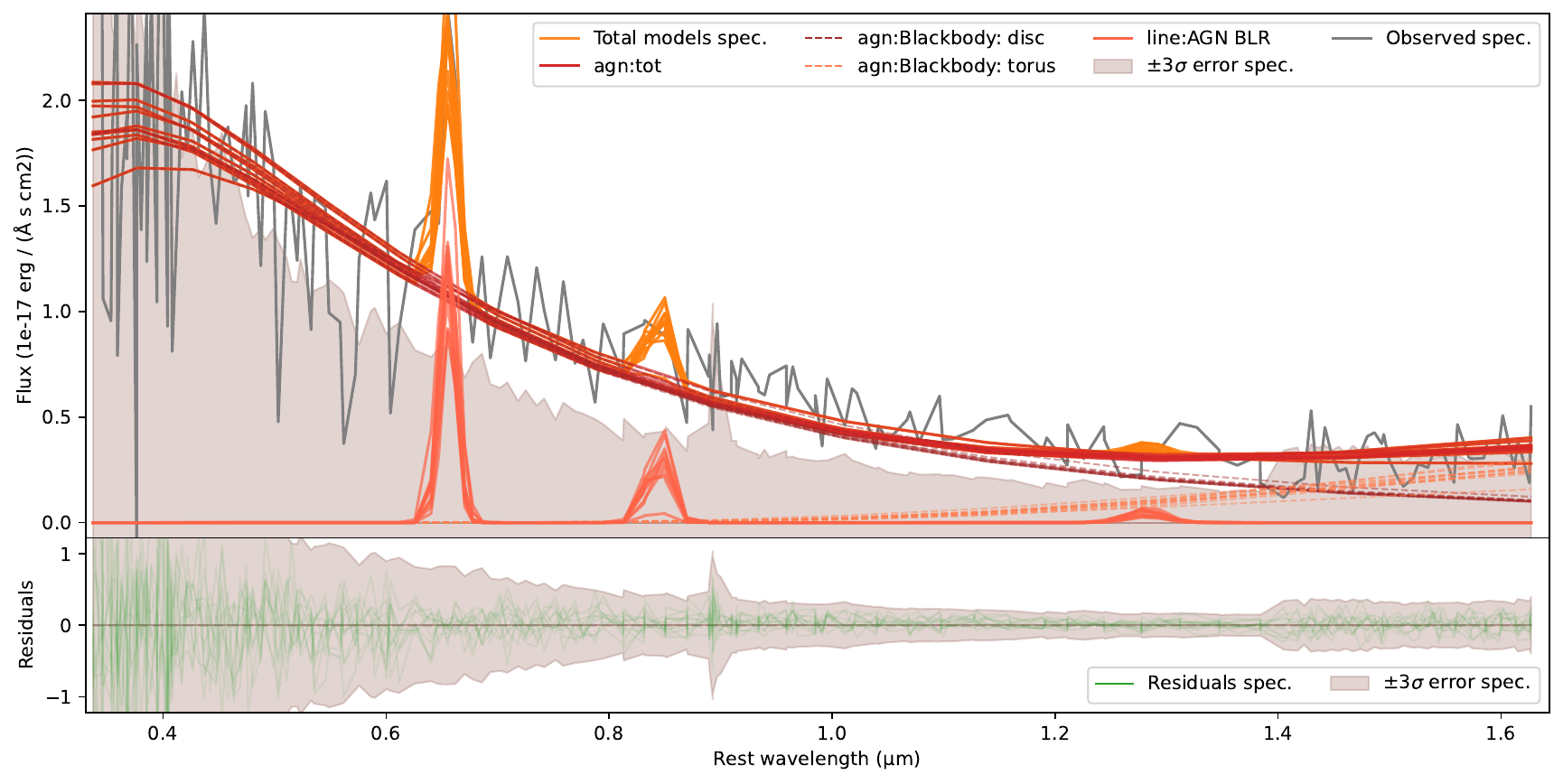}
\caption{
SPHEREx spectrophotometry (gray solid line) fitted by $\sanfit$.
From left to right, peaks of the modeled line components indicate $\halpha$, \oi, and $\mathrm{Pa\beta}$.
}\label{fig:spherex}
\end{center}
\end{figure*}

We show the SPHEREx \citep{boc26} spectrophotometry collected from Quick Release 2 \citep{spherex_qr2}, covering 0.4--1.6~\micron\ in the rest-frame, fitted by $\sanfit$ in Figure~\ref{fig:spherex}.
The underlying continua, as revealed in Figure~\ref{fig:SED}, are assumed to be a composition of SMBH envelope and dust torus blackbody components.
This spectrophotometry exhibits a robust detection of $\halpha$, while detections of \oi and $\mathrm{Pa\beta}$ are only marginal.
Due to the low resolution, it is not feasible in this stage to determine whether there exist absorption features associated with Balmer series, and the FWHM of $\halpha$ cannot be reliably measured.

Current fits find $\mathrm{FWHM(\halpha)\approx2100\pm1800\ \kms}$ and $\tnr{F}{\halpha}=(5.98\pm1.53)\times10^{15}\ \mathrm{erg/s/cm^2}$, resulting in $\tnr{L}{\halpha}=(1.25\pm0.33)\times10^{44}\ \ergs$ after correcting for the resolution.
With these measurements, we estimated the SMBH mass following \citep{shen11_smbh_mass}
\begin{equation}
    \log\left(\frac{M_{\mathrm{BH,\halpha}}}{M_\odot}\right)
= 0.379
+ 0.43 \log\left(\frac{\tnr{L}{\halpha}}{10^{42}\,\ergs}\right)
+ 2.1 \log\left(\frac{\mathrm{FWHM}_{\halpha}}{\kms}\right).
\end{equation}
Resultantly, we find $\tnr{M}{BH,\halpha}=1.8^{+5.6}_{-1.8}\times10^{8}\ \msun$, in loose agreement with $\tnr{M}{BH,MgII}=3.9^{+10.0}_{-2.8}\times10^8\ \msun$.
Nonetheless, these values should be handled with caution since we simply treat the observed $\halpha$ as the broad component originating from the BLR. 
High-resolution spectroscopy is mandatory to confirm these numbers.

%------------------fig: LRD selection misses BB-----------------%

\bibliographystyle{aasjournal}
\bibliography{BBQSO}
%\end{thebibliography}


\end{CJK*}
\end{document}